# Enhanced gas sensing performance and all-electrical room temperature operation enabled by a WSe$_2$/MoS$_2$ heterojunction

*Sushovan Dhara, Himani Jawa, Sayantan Ghosh, Abin Varghese and Saurabh Lodha*

Department of Electrical Engineering
Indian Institute of Technology Bombay
Mumbai 400076, India
E-mail: slodha@ee.iitb.ac.in

Keywords: van der Waals materials, WSe$_2$, MoS$_2$, heterojunction, gas sensor

**Abstract**

Gas sensors built using two-dimensional (2D) MoS$_2$ have conventionally relied on a change in field-effect-transistor (FET) channel resistance or a change in Schottky contact/pn homojunction barrier. This report demonstrates, for the first time, an NO$_2$ gas sensor that leverages a gate tunable type II WSe$_2$ (p)/MoS$_2$ (n) heterojunction to realize a 4x enhancement in sensitivity, 8x lower limit of detection and improved dynamic response when compared to an MoS$_2$ FET sensor on the same flake. Comprehensive sensing measurements over a range of analyte concentrations, gate biases and MoS$_2$ flake thicknesses indicate a novel two-fold electrical response to NO$_2$ exposure underlying the enhanced sensitivity of the heterojunction- (i) a series resistance change that leads to an exponential change in thermionic current at high bias, and, (ii) a carrier concentration change that leads to a linear change in interlayer recombination current near zero bias. The heterojunction diode also exhibits fast and tunable recovery under negative gate biasing. All-electrical (gate controlled) sensing and recovery operation at room temperature makes this a simple, low-overhead sensor. The ability to sense tri-nitro toluene (TNT) molecules down to a concentration of 80PPB highlights its potential as a comprehensive chemical sensing platform.

# 1. Introduction

Two-dimensional (2D) materials such as graphene and semiconducting transition metal dichalcogenides (TMDCs) are being explored extensively for chemical sensing [1,2] due to their large surface area-to-volume ratio that enables high sensitivity detection. They also offer the possibility of low power, high reliability room temperature (RT) detection[3,4] with varying affinity for different gases that ensures high selectivity.[5] This is in contrast to the poor selectivity and high operating temperatures, typically in the range of 200ºC-400ºC, seen in conventional metal oxide semiconductor based gas sensors[6] that lead to high power consumption, poor reliability and short product lifespan. Amongst the numerous TMDC semiconductors, $MoS_2$ has shown significant promise for gas sensing owing to its ambient and chemical stability,[7–10] varying affinity for different gases[11] and excellent electronic properties. [12] Exfoliated or chemical vapor deposition (CVD)-grown $MoS_2$ flakes tend to have sulphur vacancies on the surface,[13] which act as ideal adsorption sites for chemical species. $MoS_2$ exhibits intrinsic n-type transport behaviour with high electron mobilities.[14] Due to its thin conduction channel and ability to interact with the surrounding medium, electron transport in few-layer $MoS_2$ gets easily influenced on ambient exposure.[15] Exploiting this phenomenon, different $MoS_2$ based humidity and gas sensors capable of sensing moisture,[16] nitric oxide NO,[17] $NO_2$,[18] $NH_3$,[19] etc. have already been demonstrated.

Field effect transistors (FETs) with $MoS_2$ channels,[17], metal/$MoS_2$ Schottky diodes[20] and $MoS_2$ pn homojunctions have been used for $MoS_2$ chemical sensing. While the FETs respond through a change in channel conductance, Schottky diodes and pn homojunctions have been shown to work based on the modulation of the junction potential barrier resulting in higher sensitivities. However, Schottky diodes and pn homojunctions employ a single TMDC for sensing and do not offer significant flexibility in tailoring junction interface properties. $MoS_2$ sensors have also been shown to recover with the aid of optical illumination [21] or heating,[22] both of which require additional circuitry overhead. In this regard, a 2D-2D heterojunction brings forth the advantages of tailoring interface properties [23] for high sensitivity, selective dual gas detection by the two different 2D semiconductors, and all-electrical operation for both, sensing and recovery, through gate-induced selective desorption. The few-layer $WSe_2$ (p)/$MoS_2$ (n) type II heterojunction is stable in ambient conditions and works as an excellent pn diode with large rectification ratios (RR) and high degree of gate tunability.[24,25] In comparison to $WSe_2$, $MoS_2$ also has a substantially larger affinity for $NO_2$,[26,27] and hence the $WSe_2$/$MoS_2$ pn heterojunction is a good choice to demonstrate a high sensitivity, $MoS_2$ based $NO_2$ sensor with all-electrical operation.

In this work we report, for the first time, a gate tunable few layer $WSe_2$/$MoS_2$ heterostructure pn diode gas sensor with $MoS_2$ as the sensing layer and all-electrical, room temperature operation. $NO_2$ has been used as the analyte gas due to the strong electron withdrawing nature of the $NO_2$ molecule[28] that reduces free electron concentration in $MoS_2$, akin to a lowering of its donor doping.[29] The heterostructure diode is shown to consistently outperform a conventional $MoS_2$ FET sensor, fabricated on the same flake as the diode, across key sensor performance metrics such as sensitivity, limit of detection (LOD) and dynamic response. In contrast to the FET, the diode shows a two-fold electrical response to $NO_2$ exposure- (i) a series resistance change that leads to an exponential change in thermionic current at high bias, and (ii) a carrier concentration change that leads to a linear change in interlayer recombination current near zero bias, leading to > 4x overall enhancement in sensitivity. This mechanism is able to model comprehensive measurements over a range of analyte concentrations, gate biases and $MoS_2$ flake thicknesses. Besides sensing, the heterojunction diode shows excellent dynamic recovery under negative gate biasing. Room temperature operation that relies only on gate bias to alternate between sensing and recovering modes makes this a simple, low-overhead sensor. Finally, the enhanced sensitivity of the diode was exploited to sense tri-nitro toluene

(TNT) molecules down to a concentration of 80PPB, highlighting its potential as a comprehensive chemical sensing platform.

## 2. Device fabrication and electrical characterization

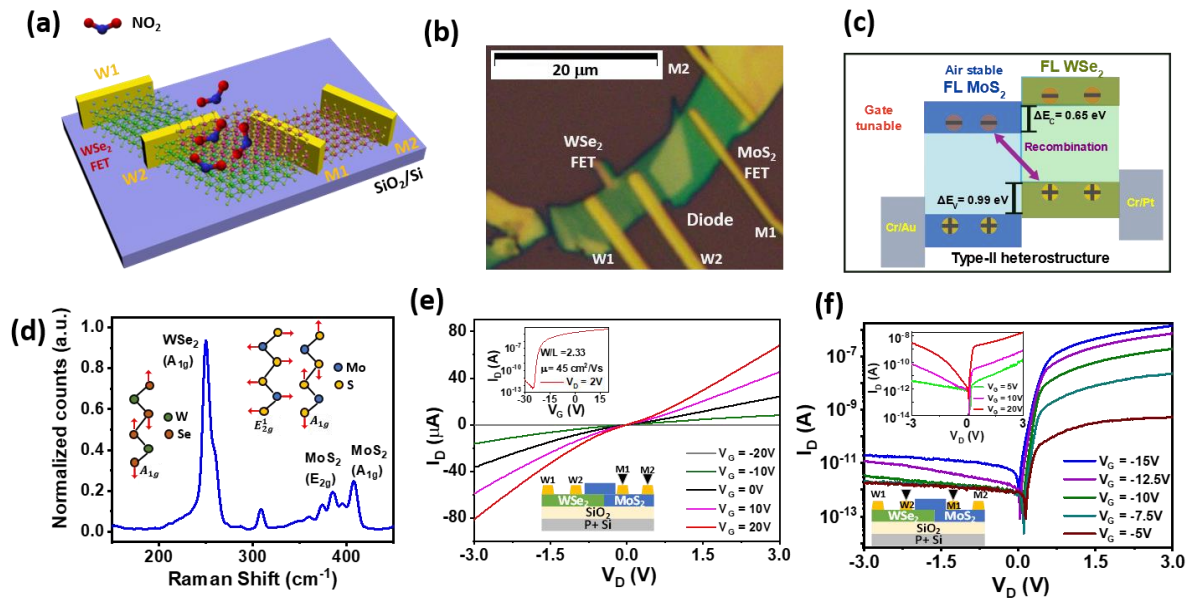

**Figure 1:** (a) Device schematics of the $WSe_2/MoS_2$ pn heterostructure and the $MoS_2$ FET, (b) optical micrograph of the fabricated diode and FET, (c) energy band diagram of the $WSe_2/MoS_2$ type-II heterostructure pn diode, (d) Raman spectrum from the heterointerface, (e) $I_D$-$V_D$ output characteristics of the $MoS_2$ FET for varying $V_G$ (inset: $I_D$-$V_G$ transfer characteristic of the FET at $V_D$=2V), and, (f) $I_D$-$V_D$ pn (negative $V_G$) output characteristics of the $WSe_2/MoS_2$ heterostructure (inset: $I_D$-$V_D$ $nn^+$ (positive $V_G$) output characteristics of the heterostructure).

The schematic shown in Figure 1(a) depicts the $WSe_2$/ $MoS_2$ heterostructure and the $MoS_2$ FET fabricated on the same $MoS_2$ flake to ensure that performance comparison between the two devices is not affected by flake-to-flake variations. Figure 1(b) shows an optical micrograph of the final fabricated diode and FET devices. For device fabrication, $WSe_2$ flakes were first exfoliated from a bulk crystal using scotch tape onto a $SiO_2/p^+$-Si substrate. $MoS_2$ flakes were then exfoliated directly onto a PDMS stamp and a selected few-layer flake was transferred on top of a selected few-layer WSe$_2$ flake using a PDMS-assisted dry transfer method.[30] Atomic force microscope scans in Supporting Information, S1, confirmed the MoS$_2$ and WSe$_2$ flakes to be few-layer thick with thicknesses of 10nm and 6nm respectively. MoS$_2$ and WSe$_2$ FET source-drain electrodes and WSe$_2$/MoS$_2$ diode contacts were patterned using electron beam lithography (EBL), and metal stacks of Cr/Au (4nm/100nm) and Cr/Pt/Au (4nm/40nm/100nm) were deposited using sputtering on MoS$_2$[31] and WSe$_2$[32] respectively to form ohmic contacts. The p$^+$-Si substrate was used as the gate electrode with the 285nm thick SiO$_2$ acting as the gate dielectric. MoS$_2$ stacked on top of WSe$_2$ forms a type II heterostructure[33] as illustrated through an energy band diagram in Figure 1(c). Raman spectrum from the heterointerface is shown in Figure 1(d). Peaks at 384cm$^{-1}$ and 407cm$^{-1}$ correspond to in-plane and out-of-plane resonance modes of MoS$_2$ respectively,[34] and the peak at 250cm$^{-1}$ corresponds to the in-plane resonance mode of WSe$_2$.[35] The presence of Raman peaks corresponding to MoS$_2$ as well as WSe$_2$ confirms formation of the WSe$_2$/MoS$_2$ heterointerface. The MoS$_2$ and WSe$_2$ FETs were characterized independently by probing contacts M1, M2 and W1, W2 (Figure 1a) respectively. As expected, both FETs showed gate biasing ($V_G$) dependent drain currents ($I_D$). The $I_D$-$V_D$ output characteristics of the MoS$_2$ FET in Figure 1(e) and of the WSe$_2$ FET in SI confirm ohmic contact formation to the two flakes. $I_D$-$V_G$ transfer characteristic in Figure 1(e) inset shows n-type transport in the MoS$_2$ FET with a field-effect mobility of 46 cm$^2$/Vs and an $I_{on}$/$I_{off}$ ratio of 10$^8$. On the other hand, transfer characteristics of the WSe$_2$ FET in Supporting Information, S2.b indicate ambipolar transport behaviour. Hence, for the diode, there are two distinguishable regions of operation. Contact M1 on the MoS$_2$ flake and contact W2 on the WSe$_2$ flake were used to characterize the diode. When a negative bias is applied to the gate, the MoS$_2$ flake shows n-type behaviour, whereas the WSe$_2$ flake exhibits p-type conduction, resulting in a pn diode configuration for the heterostructure. However, when positive gate bias is applied to the substrate, the MoS$_2$ flake becomes n$^+$ whereas WSe$_2$ becomes n-type, resulting in an nn$^+$ configuration. Figure 1(f) and its inset show the $I_D$-$V_D$ characteristics of the pn and $nn^+$ diode respectively. The pn diode shows a RR of $7.5 \times 10^4$ at gate biasing of -15V. In comparison, the $nn^+$ diode has a much lower RR of 6.5 at gate biasing of 20V. Sensing performance of both pn and $nn^+$ configurations has been measured and analysed in this work and described in detail later in this manuscript. Further, the gate bias is shown to play a crucial role in recovery of the device after sensing.

## 3. Electrical characterization with NO$_2$ exposure

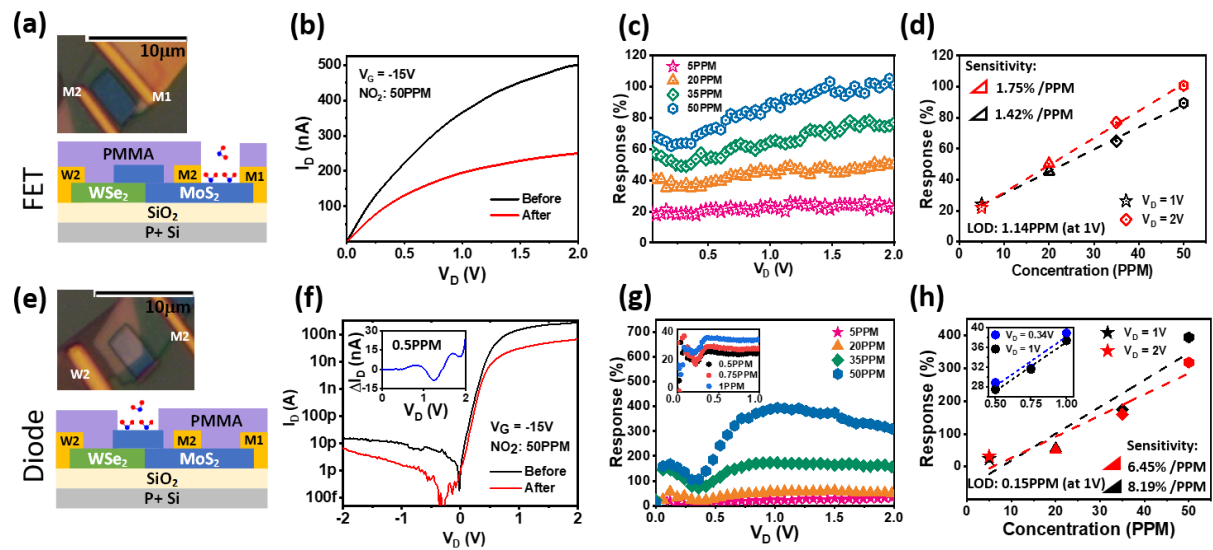

**Figure 2:** Device schematics and optical micrographs depicting NO$_2$ sensing windows opened using a PMMA mask exclusively over (a) the MoS$_2$ FET, and (e) the heterointerface overlap region of the WSe$_2$/MoS$_2$ diode. Output characteristics ($I_D$-$V_D$) of (b) the MoS$_2$ FET, and, (f) the WSe$_2$/MoS$_2$ diode (inset: change in $I_D$ vs $V_D$ at 0.5PPM), before and after 50PPM NO$_2$ exposure showing reduction in drain current. Sensing response vs $V_D$ for varying NO$_2$ concentrations for (c) the MoS$_2$ FET, and, (g) the WSe$_2$/MoS$_2$ diode (inset: response for sub-5PPM NO$_2$ concentrations). Sensing response vs NO$_2$ concentration ($V_D$ = 1V and 2V; $V_G$ = -15V) for (d) the MoS$_2$ FET, and, (h) the WSe$_2$/MoS$_2$ diode (inset: response for low NO$_2$ concentrations). Extracted LOD and sensitivity values for the FET and the diode are also shown in (d) and (h) respectively. The calculation of LOD is discussed in Supporting Information, Table S1.

Sensing measurements for the FET and the diode were independently performed in a sequential fashion but under identical conditions to ensure a fair performance comparison. For the FET sensing experiment, the entire sample was first spin coated with 1.6μm thick e-beam resist (PMMA 950K 4%) and then selectively removed only from the top of the $MoS_2$ FET channel using EBL. Hence, the $MoS_2$ FET channel was exclusively exposed to the ambient while the rest of the sample was covered under PMMA as shown in Figure 2(a). Similarly for the diode sensing experiment, the previous PMMA layer was removed, the entire sample was re-coated with PMMA and a sensing window, with the same area (6.5um x 3um) as in case of the FET, was opened on top of the $WSe_2/MoS_2$ heterointerface by EBL as shown in Figure 2(e). The sample was wire-bonded to a chip carrier for the sensing measurements, ensuring that the FET and diode currents and ohmic nature of their contacts are not affected.

For the sensing measurements, 1000PPM of $NO_2$ with balance $N_2$ was mixed with 5N $N_2$ to obtain $NO_2$ with concentrations ranging from 0.5PPM to 100PPM by controlling the flow rates of the stock gases using mass flow controllers. The measurements were conducted inside a closed chamber using a Linkam® stage under dark conditions at atmospheric pressure and room temperature. The sensing measurement setup is described in detail in SI. The FET and diode were probed using the Linkam® stage that was cyclically vacuumed and flushed ten times with 5N grade $N_2$ ten times to minimize the presence of other gas molecules inside the chamber. The sample was initially kept under 200 sccm of $N_2$ flow for 30 minutes; this helped in obtaining a stable baseline $I_D$-$V_D$ and in evaluating the noise performance of the measurement setup. $NO_2$ with varying concentration (0.5PPM, 0.75PPM, 1 PPM, 5PPM, 20PPM, 35PPM, and 50PPM) at 200 sccm was passed through the chamber for 10 minutes. The devices (FET/diode) were characterized electrically at each concentration. Before every individual exposure of $NO_2$, the device current was restored to its initial condition ($I_D$-$V_D$ characteristics) using gate voltage pulsing, which is discussed in a later section.

Figures 2(b) and 2(f) show reduction in $I_D$ with varying $V_D$ for the FET and the diode respectively after a ten minute exposure to a fixed concentration of 50PPM. Unlike the FET that could not sense below 5PPM, the diode showed a significant change in $I_D$ ($\Delta I_D$) down to the lowest concentration value of 0.5PPM (inset of Figure 2(f)). The FET showed a reduction in $I_D$ for $NO_2$ concentrations of only 5PPM and above. The response of the FET is consistent with literature.[36] Since $NO_2$ is a strong electron-withdrawing group, adsorbed $NO_2$ molecules on the surface of the $MoS_2$ FET channel reduce channel electron concentration leading to a reduction in $I_D$. The diode, on the other hand, shows a reduction in forward as well as reverse bias. The mechanism for diode response is discussed in detail later in the manuscript.

Sensing response is calculated as $|(I_{D\_after} - I_{D\_before})/I_{D\_after}|$, where $I_{D\_before}$ is the before-exposure drain current and $I_{D\_after}$ is the after-exposure drain current. Figures 2(c) and 2(g) depict the $V_D$-dependent sensing response of the FET and the diode respectively. The diode shows maximum sensing response near 0.75V forward bias at the transition from ideal to series resistance-limited transport, whereas the FET shows maximum response at maximum $V_D$ of 2V. Further, the diode shows higher response than the FET consistently over the entire $V_D$ range and a larger concentration range (inset of Figure 2(g)). Figures 2(d) and 2(h) show the calibration curves (sensing response vs concentration) of the FET and diode respectively. The inset of Figure 2(h) shows the sub-1PPM calibration curve for the diode. Slope of the calibration curve gives the device sensitivity. At $V_D$ of 1V as well as 2V the diode shows an 3.5-5.5x enhancement in sensitivity over the FET- 1.42%/PPM vs 6.45%/PPM and 1.75%/PPM vs 8.19%/PPM respectively. The limit of detection (LOD) was calculated by following the standard IUPAC procedure as described in SI.[37] The $WSe_2/MoS_2$ diode has an LOD of 0.15PPM, nearly 8 times lower than 1.14PPM of the $MoS_2$ FET fabricated using the same $MoS_2$ flake. Thus, from an overall sensing perspective, the diode has an advantage over the

FET in terms of key sensing metrics such as sensing response, sensitivity and LOD, as well as power consumption due to potential low voltage operation.

The WSe$_2$/MoS$_2$ diode current and MoS$_2$ FET drain current, at +20V and -15V gate biasing for concentrations of 0.5PPM, 0.75PPM, 5PPM, 35PPM and 50PPM are shown in Supporting Information, S3–S4 and S5–S6 respectively.

## 4. Sensing mechanism of heterojunction diode

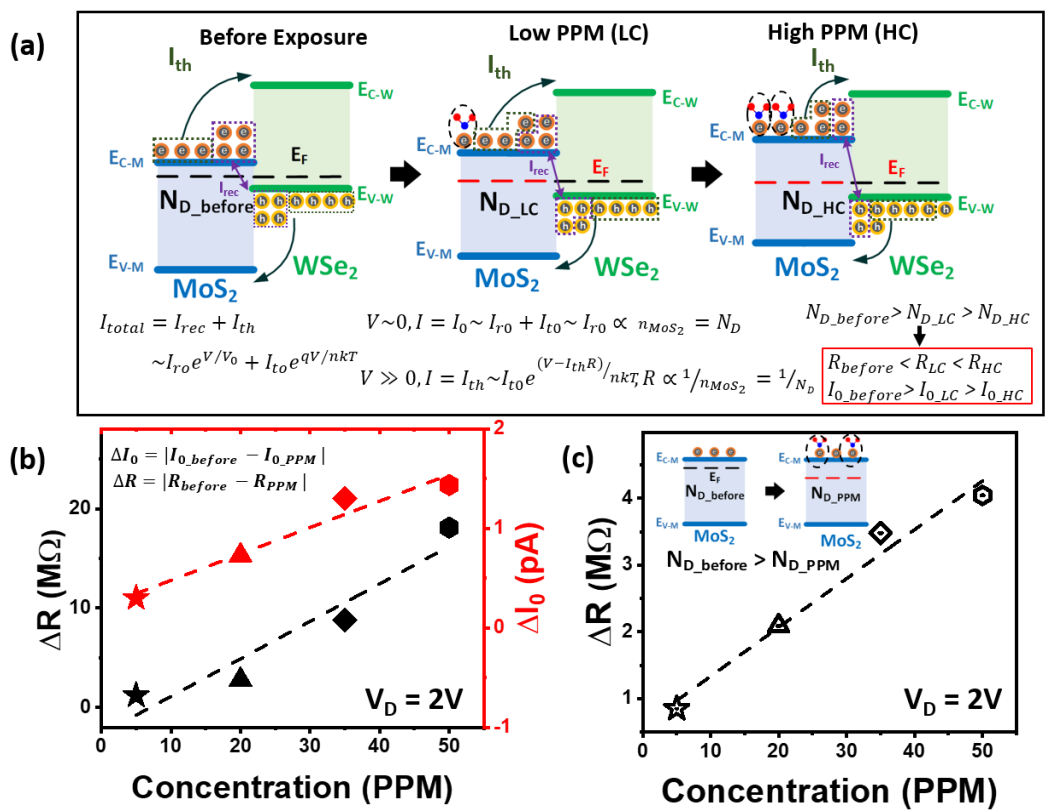

**Figure 3:** (a) Equilibrium band diagrams of the WSe$_2$/MoS$_2$ diode before and after low PPM and high PPM NO$_2$ exposure. With increasing NO$_2$ concentration the effective doping in the MoS$_2$ flake reduces leading to a decrease in low-bias interlayer recombination current as well as high-bias thermionic current due to increased series resistance. (b) Extracted change in I$_0$ (ΔI$_0$) and series resistance (ΔR) of the diode for varying NO$_2$ concentration. (c) Change in channel resistance (ΔR) of the MoS$_2$ FET for varying NO$_2$ concentration.

In reported Schottky diode[38] and pn homojunction[39] sensors, adsorption of NO$_2$ changes the effective doping of the sensing layer leading to a shift in its Fermi level and a change in the junction built-in potential.[40] The series resistance also increases with NO$_2$ adsorption,[21] which further reduces the device current. However, in the case of van der Waals pn heterojunctions such as the WSe$_2$/MoS$_2$ heterostructure in this study, in addition to the thermionic current (I$_{th}$), the interlayer recombination current (I$_{rec}$) has also been shown to play a significant role in current transport.[41,42] As shown in Figure 3(a) the total diode current (I$_{total}$) can be expressed as $I_{total} = I_{rec} + I_{th} \sim I_{r0}e^{V/V_0} + I_{t0}e^{qV/nkT}$,[43] where V is the applied bias, I$_{r0}$ and I$_{t0}$ are the saturation currents for the recombination and thermionic contributions, q is unit electron charge, k is Boltzmann's constant, n is ideality factor, V$_0$ is a constant and T is temperature. I$_{rec}$ is similar to the excess current seen in p$^+$/n$^+$ Esaki diodes.[43] Near zero bias, the total saturation current, I$_0$, is dominated by I$_{r0}$ which in turn depends on the product of electron concentration in MoS$_2$ ($n_{MoS_2}$) and hole concentration in WSe$_2$ ($p_{WSe_2}$) whether it is band-to-band or trap-assisted recombination.[42] At high bias, I$_{th}$ dominates and with increasing current density, the I$_{th}$R voltage drop due to series resistance R becomes significant, where R is inversely proportional to $n_{MoS_2}$. In summary, the diode current depends strongly on $n_{MoS_2}$ through recombination-dominated near zero bias saturation current as well series resistance-limited high bias thermionic current. As illustrated through equilibrium energy band diagrams in Figure 3(a), increasing NO$_2$ concentration from before exposure → low PPM (LC) → high PPM (HC), progressively decreases $n_{MoS_2}$ by immobilizing conduction band electrons and hence the effective MoS$_2$ doping, N$_D$. This leads to decreasing recombination (I$_{0\_before}$ > I$_{0\_LC}$ > I$_{0\_HC}$) as well as thermionic current (increasing R → R$_{before}$ < R$_{LC}$ < R$_{HC}$). This two-fold response of the diode offers the benefits of (i) enhanced sensing response and sensitivity if operated at high-bias due to the exponential dependence of I$_{th}$ on R, and, (ii) reduced power consumption if operated near zero-bias but with a reduced sensing response and sensitivity because of a linear dependence of I$_0$ on $n_{MoS_2}$. It is important to note that only the top MoS$_2$ flake is exposed to $NO_2$ at the heterointerface and hence $p_{WSe_2}$, and its contributions to $I_{r0}$ and R, remain unchanged with $NO_2$ exposure.

Figure 3(b) shows the extracted change in R (ΔR) and $I_0$ (Δ$I_0$) for the diode with increasing $NO_2$ concentration. As expected from the mechanism described in Figure 3(a), ΔR and Δ$I_0$ change linearly with $NO_2$ concentration. R increased by 18MΩ and $I_0$ decreased by 1.1pA when the concentration was increased from 5 to 50PPM. Figure 3(c) shows the extracted change in channel resistance, ΔR, for the control $MoS_2$ FET for increasing $NO_2$ concentration at $V_G$=-15V and $V_D$=2V. As expected, ΔR increases linearly by about 3MΩ when $NO_2$ concentration was increased from 5 to 50PPM, and since the drain current has a linear dependence on the channel resistance,[44] the sensing response of the FET is substantially lower than the diode.

## 5. Gate biasing dependent sensing performance

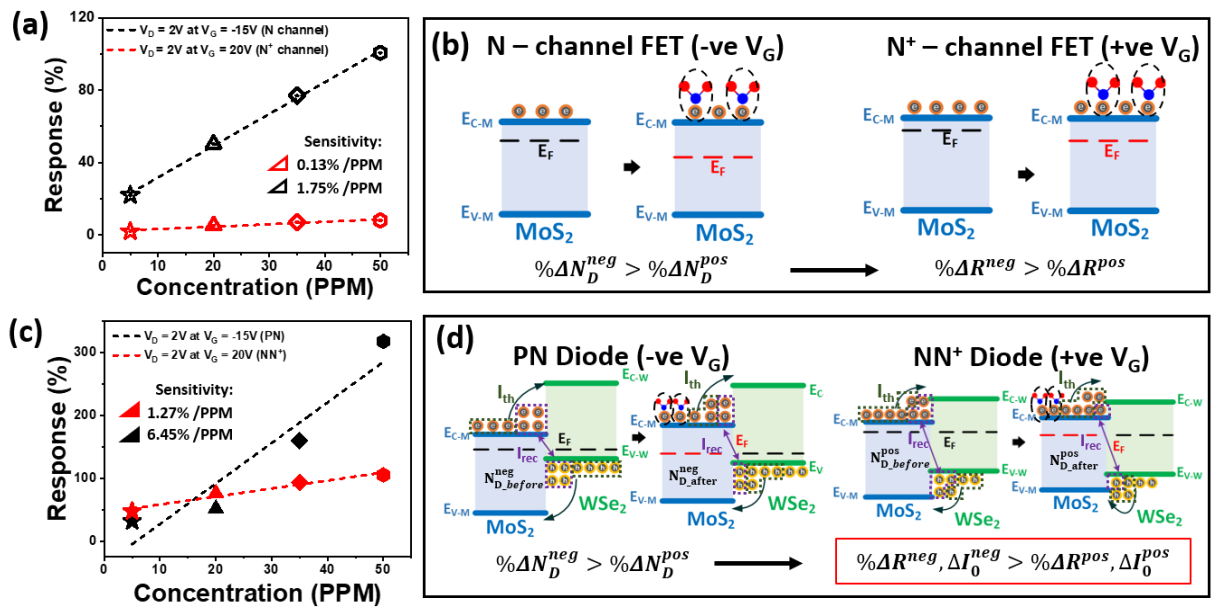

**Figure 4:** Positive ($n^+$ channel) and negative (n channel) gate bias dependence of (a) sensing response versus $NO_2$ concentration and sensitivity, and, (b) before and after $NO_2$ exposure energy band diagrams of control $MoS_2$ FET. Positive ($nn^+$ diode) and negative (pn diode) gate bias dependence of (c) sensing response versus $NO_2$ concentration and sensitivity, and, (d) before and after $NO_2$ exposure energy band diagrams of $WSe_2/MoS_2$ heterostructure.

Next, the gate bias dependent sensing performance of the FET and the diode was examined for $V_G=-15V$ and $V_G=+20V$. At -15V of gate biasing, the $MoS_2$ FET has an n-type channel, and the diode is in pn configuration, whereas at +20V of gate voltage, the FET has an $n^+$-channel, and the diode is in $nn^+$-configuration. Figure 4(a) shows the sensing response of n- and and $n^+$-channel $MoS_2$ FET for $NO_2$ concentration ranging from 5PPM to 50PPM. The n-channel configuration has a higher response as well as sensitivity of 1.75%/PPM compared to the $n^+$ configuration (0.13%/PPM). This is explained through band diagrams in Figure 4(b). The $n^+$-channel has higher electron concentration than the n-channel configuration. However, the number of adsorption sites available for $NO_2$ molecules are the same in both cases. Hence, for a given $NO_2$ concentration although the number of electrons immobilized in both channels is the same, the percentage change in effective doping as well as the channel resistance (%ΔR) is less in the $n^+$-channel case. Similarly, Figure 4(c) shows that the pn configuration has better sensing response and sensitivity (6.45%/PPM vs 1.27%/PPM) compared to the $nn^+$ case for the $WSe_2/MoS_2$ diode. As seen from the energy band diagrams for the heterojunction in Figure 4(d), %ΔR is higher for negative $V_G$ following the same logic as in the case of FET since it is only the $MoS_2$ layer that plays a role in determining the change in R. As $V_G$ changes from negative to positive, $MoS_2$ goes from n to $n^+$ resulting in a smaller %ΔR. The pn configuration has substantial $I_{rec}$ whereas the $nn^+$ configuration behaves more like an n-channel FET, with low $I_{rec}$ due to a large interlayer gap (Figure 4(d)). Hence from a combined R and $I_0$ perspective, the pn configuration shows a higher normalized change in $I_{total}$ resulting in a better sensing response. Comparing the gate bias dependence of the diode and the FET shows that firstly, the best-case for the FET (negative $V_G$) is quantitatively similar to the worst-case for the diode (positive $V_G$) in terms of sensing response and sensitivity, and secondly, this shows that the $nn^+$ diode is effectively an n-channel FET, i.e. the diode performance can be tuned from diode → FET using gate bias.

## 6. Flake thickness dependent sensing performance

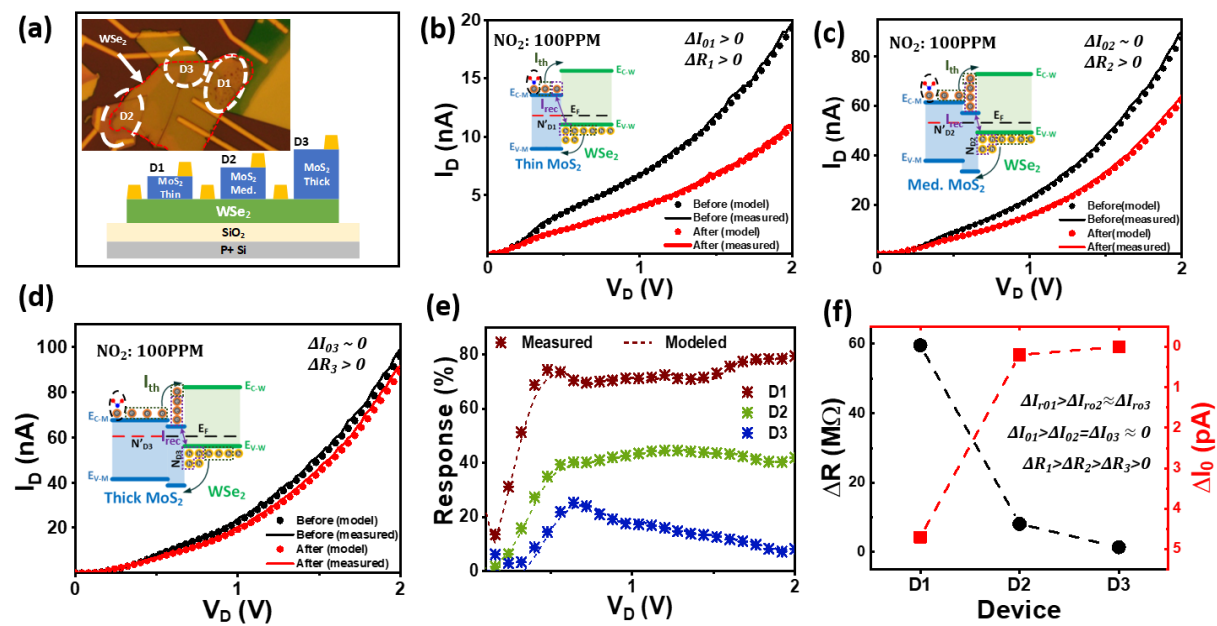

**Figure 5:** (a) Optical micrograph and schematic showing three WSe$_2$/MoS$_2$ heterojunction diodes on the same WSe$_2$ flake but with MoS$_2$ thickness increasing from thin (D1) → medium (D2) → thick (D3). Measured and modeled (Supporting Information, S10) forward bias I$_D$-V$_D$ output characteristics of (b) D1, (c) D2, and, (d) D3, before and after NO$_2$ exposure. The change in drain current decreases with increasing MoS$_2$ thickness. (e) Comparative drain voltage dependent sensing response plot for D1, D2 and D3 confirms higher response for thinner MoS$_2$ i.e. for D1 followed by D2 and D3. (f) Extracted ΔI$_0$ and ΔR values for D1, D2 and D3 decrease with increasing MoS$_2$ thickness.

Flake thickness is known to significantly influence electrical and optical properties in 2D materials. Hence, MoS$_2$ thickness dependence of the WSe$_2$/MoS$_2$ diode sensor performance was carried out to gain further insights into the sensing mechanism. MoS$_2$ flakes of three different thicknesses ~6 nm (thin), ~45 nm (medium) and 60 ~nm (thick) (Supporting Information, S7) were transferred on top of a single WSe$_2$ flake. Same as before, Cr/Au contacts on MoS$_2$ and Cr/Pt/Au contacts on WSe$_2$ were fabricated. The transfer characteristics confirmed the ambipolar nature of WSe$_2$ and n-type nature of MoS$_2$ (Supporting Information,

S8 e, f, g, h). Further, the output characteristics of the individual MoS$_2$ and WSe$_2$ FETs confirmed the ohmic nature of the device contacts (Supporting Information, S8 a, b, c, d). Figure 5(a) shows a microscope image and device schematic of the resulting three heterojunction diodes labelled as D1, D2, and D3, where D1 represents the thin MoS$_2$ flake, D2 corresponds to the MoS$_2$ flake of medium/intermediate thickness, and D3 represents the thick MoS$_2$ flake. For the sensing measurements, the entire sample was coated with PMMA and identical sensing windows were opened over each heterointerface using EBL, one at a time. All the three diodes were exposed to 100 PPM of NO$_2$ for 10 minutes. Figures 5(b), (c) and (d) show the reduction in forward bias $I_D$ of D1, D2 and D3 respectively with NO$_2$ exposure. The reduction in $I_D$ follows a clear trend, i.e. it decreases with increasing MoS$_2$ flake thickness. Figure 5(e) depicts the extracted sensing response, confirming the observation that thinner flakes result in a larger percentage change in $I_D$. As seen earlier, the sensing response peaks near the transition around 0.4V-0.5V from ideal to series resistance-limited current transport. The energy band diagrams in Figures 5(b), (c) and (d) illustrate that the MoS$_2$ thickness dependent sensing response is consistent with the mechanism described earlier. With increasing MoS$_2$ thickness in the vertical direction, away from the heterointerface, the reduction in $n_{MoS_2}$ and effective $N_D$ due to adsorbed NO$_2$ moves further away from the WSe$_2$/MoS$_2$ interface. Hence thicker flakes (D2 and D3) are expected to show only a change in series resistance ($\Delta R$) unlike a thin flake (D1) that will not only exhibit a $\Delta R$ but also a change in interlayer recombination current ($\Delta I_0$) due to a change in $n_{MoS_2}$ near the interface with WSe$_2$. Extracted $\Delta R$ and $\Delta I_0$ values for D1, D2 and D3, using the combined thermionic-recombination model described earlier are shown in Figure 5(f). The trends in $\Delta R$ and $\Delta I_0$ with MoS$_2$ flake thickness reinforce the sensing mechanism. $\Delta R$ decreases with increasing flake thickness from D1 → D2 → D3 due to fewer conduction electrons in thinner flakes resulting in a larger change in R. Further, $\Delta I_0$ is nearly zero for D2 and D3.

## 7. Dynamic sensing and recovery response

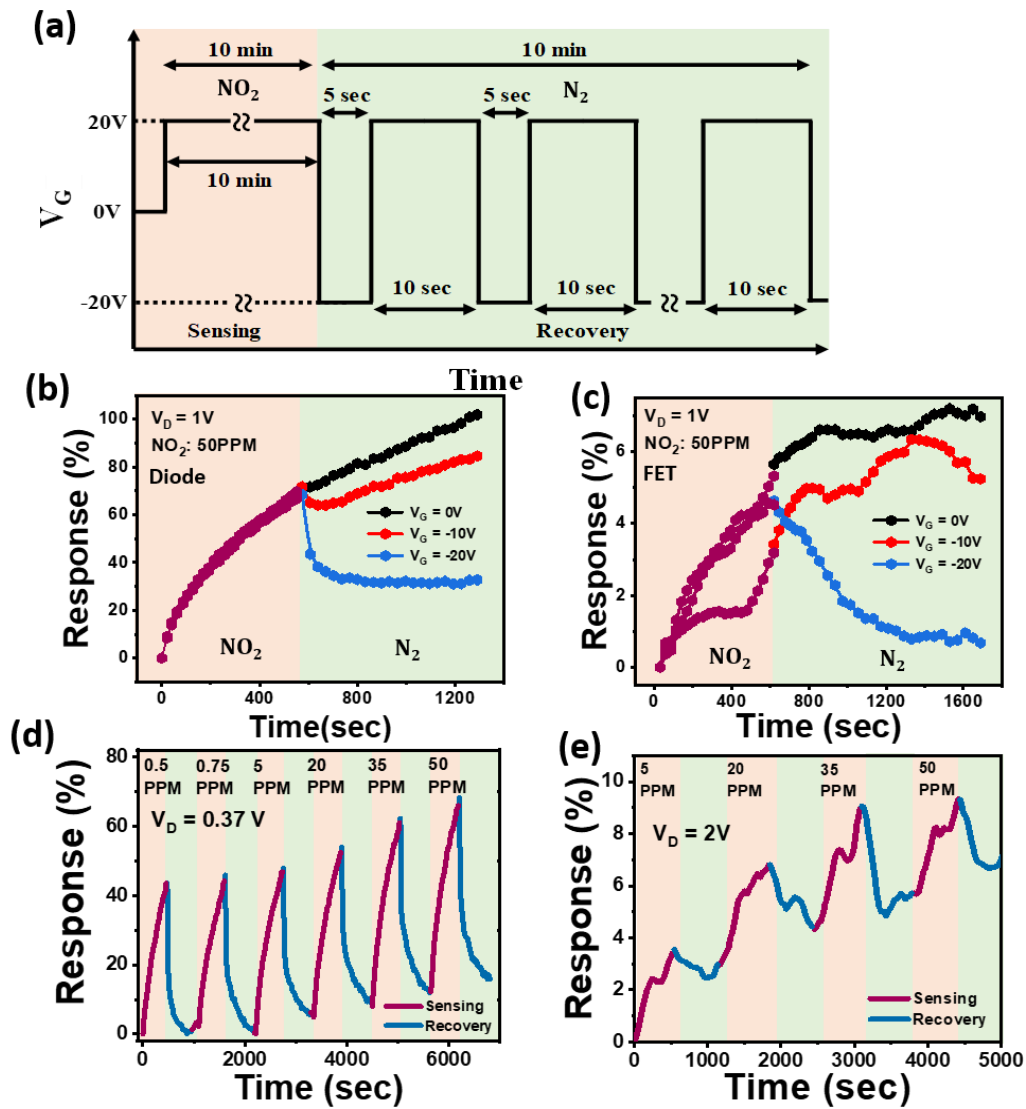

**Figure 6:** (a) Profile of applied gate voltage versus time for dynamic sensing ($V_G$=20V) and recovery ($V_G$=-20V) response measurements. Gate voltage dependence of recovery for a single time cycle of sensing ($V_G$=20V) and recovery ($V_G$=0, -10, -20V) response for (b) the WSe$_2$/MoS$_2$ diode, and, (c) the MoS$_2$ FET. Dynamic sensing ($V_G$=20V) and recovery ($V_G$=-20V) response over multiple time cycles for (d) the WSe$_2$/MoS$_2$ diode, and, (e) the MoS$_2$ FET. The drain biasing dependent sensing responce at gate voltage of 20V is discussed in figure Supporting Information, S9.

Dynamic response alternating between sensing and recovery for varying analyte concentration is an important performance test for sensors. As shown later in a benchmarking table, most reported sensors rely on heat or light to desorb the analyte gas for sensor recovery. This increases the complexity of the sensor circuitry/hardware. The ability to sense as well as recover through electrical means is therefore highly desirable. Figure 6(a) shows the time profile of the applied gate bias for sensing and recovery cycles of 10 minutes each. The sensing gate bias of +20V was chosen for the 10 min sensing cycle under continuous $NO_2$ exposure to avoid recovery artefacts that could result at negative gate biases. Although this is not the optimal sensing gate voltage for a high sensing response, it still maintains a large difference in sensing response of the diode and the FET (see Figure 4) and helps illustrate the all-electrical operation of the heterojunction sensor. During the 10 min recovery cycle under continuous flow of 5N $N_2$, the gate voltage was pulsed alternately between the recovery voltage of -20V for 5s and the measurement voltage of +20V for 10s, the time taken for one $I_D$-$V_D$ sweep. Repeated $I_D$-$V_D$ sweeps were measured at $V_G$=+20V condition for both the sensing and recovery cycles to determine the time dependent sensing and recovery response shown in Figures 6(b)-(e).

Figures 6(b) and (c) indicate that the $WSe_2$/$MoS_2$ diode as well as the $MoS_2$ FET show a strong dependence on negative gate bias for recovery after exposure to 50PPM of $NO_2$ for 10 minutes. Both the diode and the FET start showing substantial and faster recovery in sensing response as $V_G$ is decreased from 0V to -20V. The recovery is more substantial for the FET since the $MoS_2$ channel is in direct contact with the $SiO_2$ layer whereas in case of the diode the intervening $WSe_2$ layer reduces the effective gate voltage acting on the $MoS_2$ layer. Gate tunability of the recovery response, and its speed, is a significant benefit enabled by the thin nature of the 2D $MoS_2$ layer. The recovery is likely due to an electrostatic desorption effect on the negatively charged adsorbed $NO_2$.[26,27] The gate bias induced recovery is unlikely to work efficiently in pn homojunctions where the p-type layer also has strong affinity for the analyte gas. Hence, the lower affinity of the p-type $WSe_2$ layer for $NO_2$, as compared to $MoS_2$,[45] in the heterojunction diode plays a critical role in enabling its gate bias driven recovery. Figures

6(d) and (e) show the dynamic sensing and recovery response of the diode and the FET respectively over multiple time cycles for a recovery voltage of -20V and increasing $NO_2$ concentration from 0.5 to 50PPM. Both devices show an increasing sensing response for increasing $NO_2$ concentration with the diode exhibiting a better sensing response than the FET. The devices recover more than 50% with the negative gate bias. The sensing and recovery times of the diode (FET) corresponding to 90% sensing and recovery were extracted to be 385 (552) and 175 (222) seconds respectively. The diode senses and recovers faster than the FET. The dynamic and sensing performance of the devices can be improved further by making the sensing gate voltage less positive and making the recovery time longer and/or the voltage more negative. Complete recovery of diode output characteristics using negative gate bias after 20PPM of $NO_2$ exposure is shown in Supporting Information, S11.

## 8. TNT sensing experiment

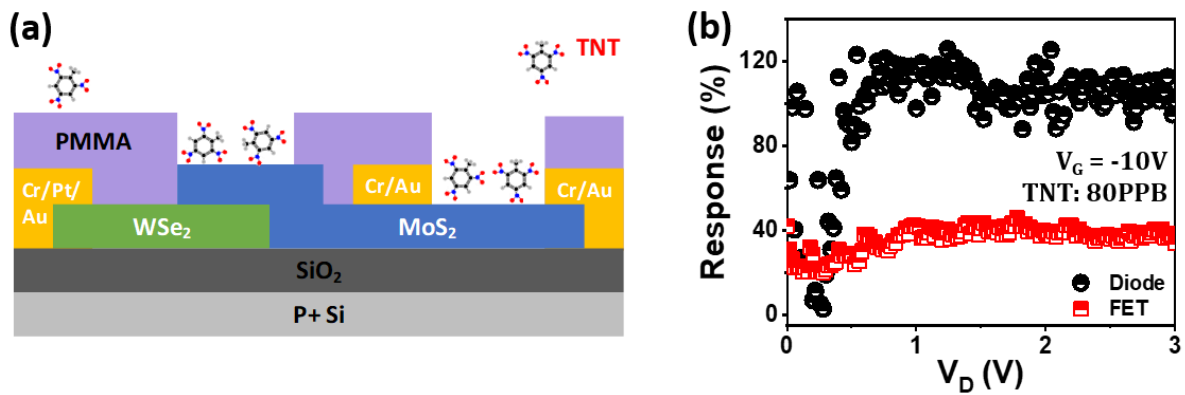

**Figure 7:** (a) Schematic showing sensing of 80PPB TNT using the $MoS_2$ FET and the $WSe_2/MoS_2$ heterointerface, (b) comparison of sensing response of the diode and FET to 80PPB TNT exposure.

Tri-nitro toluene (TNT) is an explosive gas that also contains the electron withdrawing $NO_2$ group. A similar diode-FET device as reported earlier in this manuscript was fabricated for TNT sensing (Figure 7(a)). The individual device currents have been discussed in SI. TNT has an evaporation temperature higher than room temperature. TNT vapor was generated using an Owlstone® vapor generator (details of the method have been discussed in SI) and 80PPB of TNT was obtained by mixing it with $N_2$. Both devices were simultaneously exposed to 80PPB of TNT for three minutes. The $I_D$-$V_D$ data for the FET and the diode was collected immediately after exposure at a fixed gate bias of -10V. Figure 7(b) shows a comparison of their sensing response over 3V of $V_D$. Similar to $NO_2$, the diode shows a 3x enhancement in TNT sensing response over the FET, indicating its wide applicability.

## 9. Conclusions

The table below summarizes and benchmarks key sensing and recovery parameters of this work against reported $MoS_2$ gas sensors. These include, Schottky barrier diodes (SBDs), FETs, pn homojunctions, Gr-$MoS_2$ junctions, and $MoS_2$ junctions with non-2D materials such as ZnO and $SnO_2$. In this context, the $WSe_2$/$MoS_2$ device reported in this work is the first $MoS_2$-TMDC heterojunction gas sensor. It demonstrates a unique two-fold sensing mechanism, a low bias- hence low power, low sensing response based on a change in interlayer recombination current, and a high bias- higher power consumption, high sensing response that exponentially captures the change in series resistance due to $NO_2$ adsorption. This mechanism also explains the gate bias and $MoS_2$ flake thickness dependence, where negative bias and thinner flakes show improved sensing performance. Compared to earlier reports, this is also the first device that demonstrates tunable gate bias induced recovery over a wide concentration range. The all-electrical sensing and recovery operation coupled with the ability to sense ultra-low (80PPB) concentration of TNT indicate its potential for developing a high sensitivity, low cost and reliable gas sensing platform.

**Table 1** Summary and benchmarking of key sensing and recovery parameters for $MoS_2$ gas sensors reported in literature and this work.

| Ref. | Materials | Architecture | Mechanism | Detected (% value) (ppm) | LOD (ppm) | Sensing Condition (T, L, E) | Recovery Condition (T, L, E) | Response time | Recovery time | Sensing Application |
|---|---|---|---|---|---|---|---|---|---|---|
| [20] | $MoS_2$ | SBD | SB height change | 0.02 (-) | - | RT | RT | 5 min | 12 h | $NH_3$ |
| [46] | $MoS_2$-Graphene | FET | Resistance change | 0.5 (-) | 0.014 | T | T | 1 min < | 1 min < | NA |
| [36] | 5L $MoS_2$ | FET | Resistance change | 100 (-) | - | E | RT | 3 min | 10 min | $NH_3$ and Humidity |
| [47] | Vertical $MoS_2$ | FET | Resistance change | 0.1 (0.2%) | - | RT | RT | 10 min > | 10 min > | Ethanol |
| [44] | Pt-$MoS_2$ | FET | Resistance change | 5 (-) | 0.002 | RT | RT | 30 min > | 30 min > | NA |
| [48] | $MoS_2$-Graphene | FET | Resistance change | 5 (~50%) | - | T | T | 5 min < | 10 min > | $NH_3$ |
| [3] | $SnO_2$-$MoS_2$ | FET/resistor | Resistance change | 0.5 (0.6%) | - | RT | RT | 6.8 min | 2.7 min | $H_2$, $H_2S$, CO, and $NH_3$ |
| [21] | Multi-layer $MoS_2$ | FET | Resistance change | 5 (~17%) | 0.2 | L | L | 29 sec | 350 sec | $NH_3$, $H_2$, $H_2S$, $CO_2$, and $CH_4$ |
|  |  |  |  |  |  | T | T | 71 sec | 310 sec |  |
|  |  |  |  |  |  | RT | RT | 249 sec | - |  |
| [22] | p-$MoS_2$ | FET/resistor | Resistance change | 10 (-) | - | RT | RT | 8.5 sec | Not recovered | $H_2$, $H_2S$, $CO_2$, and $NH_3$ |
|  |  |  |  |  |  | T | T | 4.44 sec | 19.6 sec |  |
|  |  |  |  |  |  | L | L | 6.09 sec | 146.49 sec |  |
| [4] | ZnO-$MoS_2$ | FET | Resistance change | 5 (3050%) | - | RT | RT | 40 sec | 300 sec | $NH_3$, $H_2S$, and $SO_2$ |
| [49] | CdTe-$MoS_2$ | PN | $V_{built-in}$ change | 10 (26%) | - | RT | RT | 16 sec | 114 sec | $NO_2$ |
| [39] | pn-$MoS_2$ | PN | $V_{built-in}$ change | 0.1 (-) | 0.15 | L | L | 60 sec | 80 sec | Trimethylamine |
| [50] | RGO $MoS_2$ | FET | Resistance change | 1 (3%) | 0.07 | RT | RT | - | - | $NH_3$, $SO_2$, CO |
| [51] | $MoS_2$ on PI | FET | Resistance change | 2 (45%) | - | RT | RT | 2.14 min | - | HCHO, $H_2$, CO, $NH_3$, and $C_2H_5OH$ |
| [52] | $MoS_2$-Gr contact | FET | Resistance change | 25 (280%) | 0.0001 | L | L | - | - | NA |
| [53] | $MoS_2$/ZnO | PN | Resistance change | 0.2 (226%) | 0.0001 | L | L | 75 sec | 11 sec | NA |
| [40] | $MoS_2$-Graphene | SBD | SB height change | 1 (-) | - | RT | RT | - | - | NA |
| **This work** | $WSe_2/MoS_2$ | PN | **Recombination current and resistance change** | 0.5 (45%) | 0.15 | E | E | 6.4 min | 2.9 min | $NO_2$, TNT |
| **This work** | $MoS_2$ | FET | **Resistance change** | 5 (3.8%) | 1.14 | E | E | 9.2 min | 3.7 min | $NO_2$, TNT |

T: Temperature, L: Light, E: Electrical

# Experimental section

*Device fabrication*

The $WSe_2/MoS_2$ heterojunction diode and the $WSe_2$ and $MoS_2$ FETs were fabricated on an $SiO_2$ (285nm)/p+ Si substrate. $MoS_2$ and $WSe_2$ flakes were exfoliated from bulk crystals (SPI Supplies) and transferred onto the substrate using scotch tape. Acetone cleaning was used to remove organic residues before and after the flake transfer. Raman spectra were collected using a LabRAM HR800 (HORIBA Scientific) system with a 532nm laser and a spot size of 1μm. MFP-3D (Asylum Research Inc.) was used for AFM measurements.

For contact fabrication, the samples were spin coated with a bilayer e-beam resist (EL9/PMMA-A4). Patterning was carried out using a Raith 150-Two direct write e-beam lithography system. Contact metals Cr/Au (for $MoS_2$) and Cr/Pt/Au (for $WSe_2$) were sputter deposited using a 7-target sputtering system from AJA International. Metal lift-off was carried out by soaking the samples overnight in acetone.

*Electrical sensing characterization*

Electrical measurements were carried out in a Linkam® stage using a Keysight B1500A parameter analyser. For $NO_2$ sensing experiments, the devices were wire bonded onto a PCB. Stock gas of 1000PPM $NO_2$/balance $N_2$ (Med Gas N Equipment) was mixed with 5N $N_2$ (Med Gas N Equipment) and the flow rates were controlled using mass-flow controllers (MKS) to obtain analyte $NO_2$ gas with varying concentration. TNT vapor was generated using an Owlstone® vapor generator employing a TNT permeation tube where 5N $N_2$ was used as the vapor carrier gas.

**Acknowledgements**

The authors H. J. and S. G. made equal contribution to the work. S.D. acknowledges the junior research fellowship (JRF) scheme from Department of Science and Technology, Govt. of India. Authors acknowledge IITB Nanofabrication Facility (IITBNF) for usage of its facilities for device fabrication and characterization. The authors acknowledge Kartikey Thakar for helping with the AFM measurements. This work was funded by the Department of Science and Technology, Govt. of India through its Swarna Jayanti fellowship scheme (Grant No. DST/SJF/ETA-01/2016-17).

**Supporting Information**

SI is included.

# Supporting Information

**Enhanced gas sensing performance and all-electrical room temperature operation enabled by a WSe₂/MoS₂ heterojunction**

*Sushovan Dhara, Himani Jawa, Sayantan Ghosh, Abin Varghese, and Saurabh Lodha*

Sushovan Dhara, Himani Jawa, Sayantan Ghosh, Abin Varghese, and Saurabh Lodha
Department of Electrical Engineering
Indian Institute of Technology Bombay
Mumbai 400076, India
E-mail: slodha@ee.iitb.ac.in

*The authors H. J. and S. G. made equal contributions to this work.*

**Table of contents**

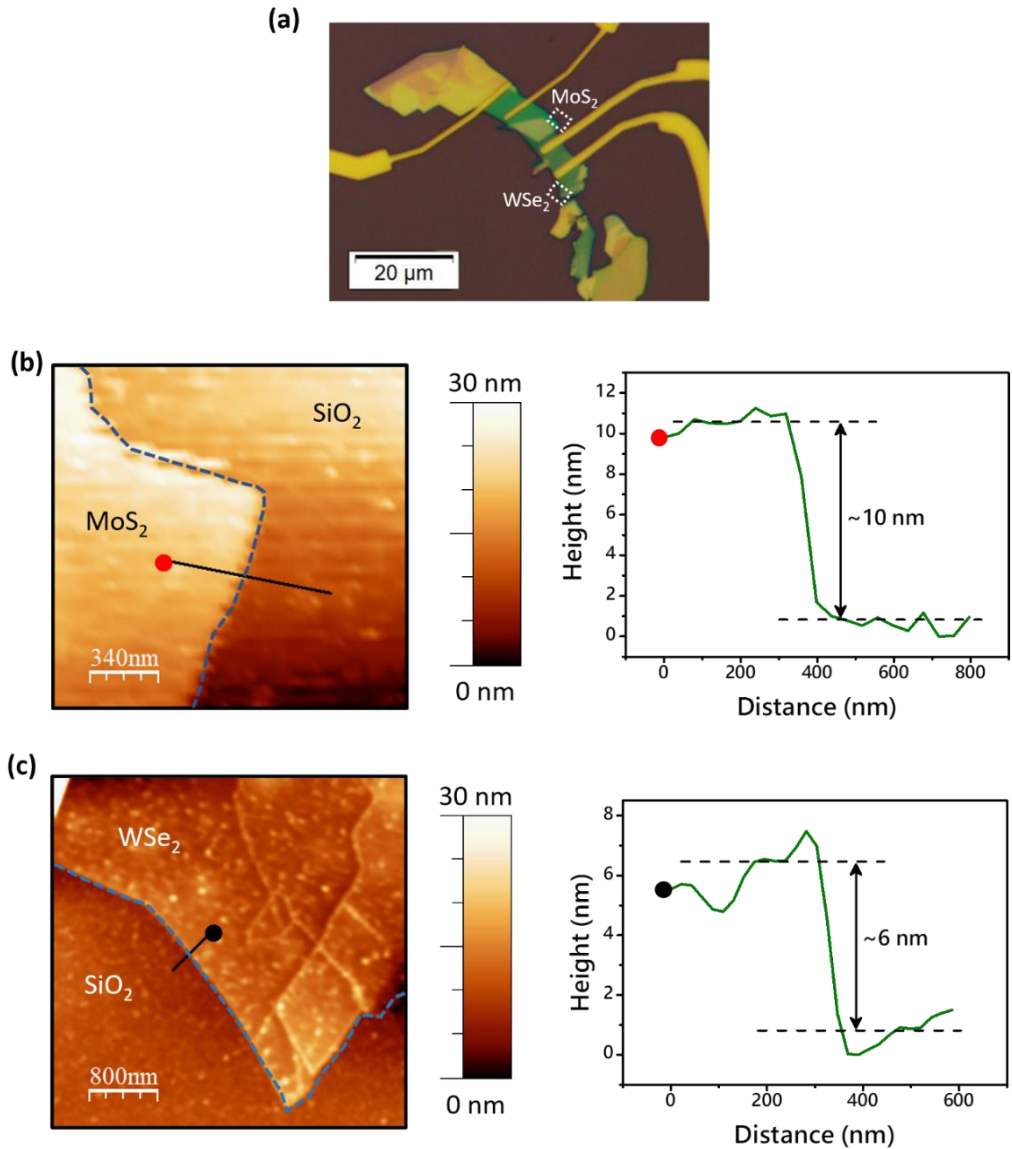

**Figure S1.** (a) Optical micrograph of the WSe$_2$/MoS$_2$ heterojunction scanned to obtain AFM images and linescans of the (b) MoS$_2$ and (c) WSe$_2$ flakes used in this work.

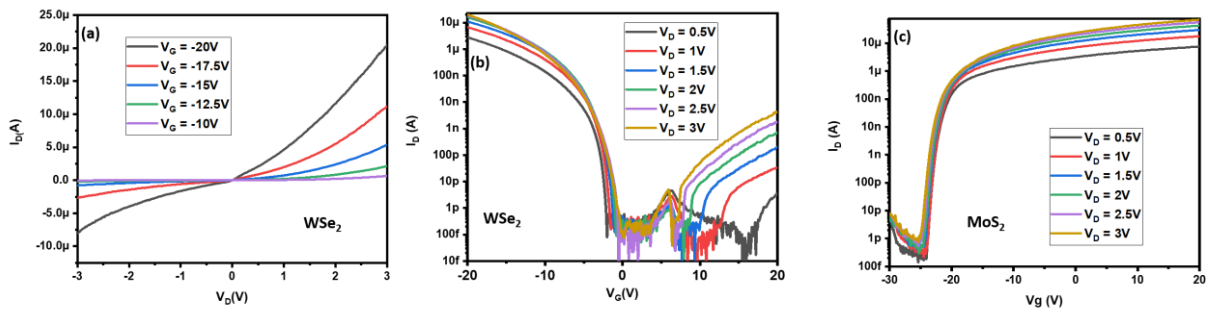

**Figure S2.** (a) $I_D$-$V_D$ output characteristics and (b) $I_D$-$V_G$ transfer characteristics of the WSe$_2$ FET confirm its ambipolar nature. (c) $I_D$-$V_G$ transfer characteristics of the MoS$_2$ FET indicate n-type behavior.

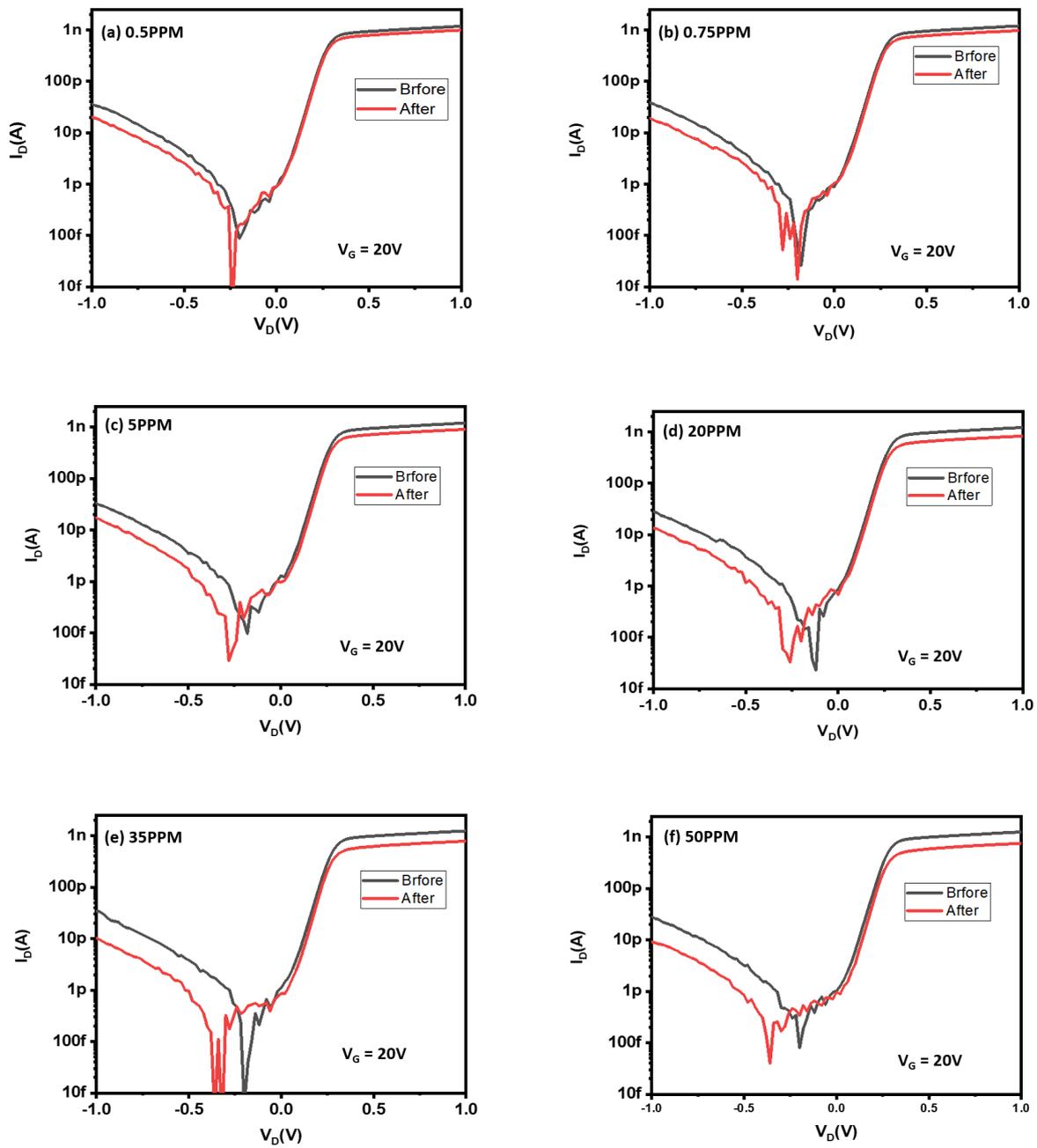

**Figure S3.** $I_D$-$V_D$ characteristics of the WSe$_2$/MoS$_2$ pn diode ($V_G$=+20V) before and after exposure to NO$_2$ gas at varying concentrations of (a) 0.5PPM, (b) 0.75PPM, (c) 5PPM, (d) 20PPM, (e) 35PPM, and, (f) 50PPM. The reduction in drain current increases with increasing NO$_2$ concentration.

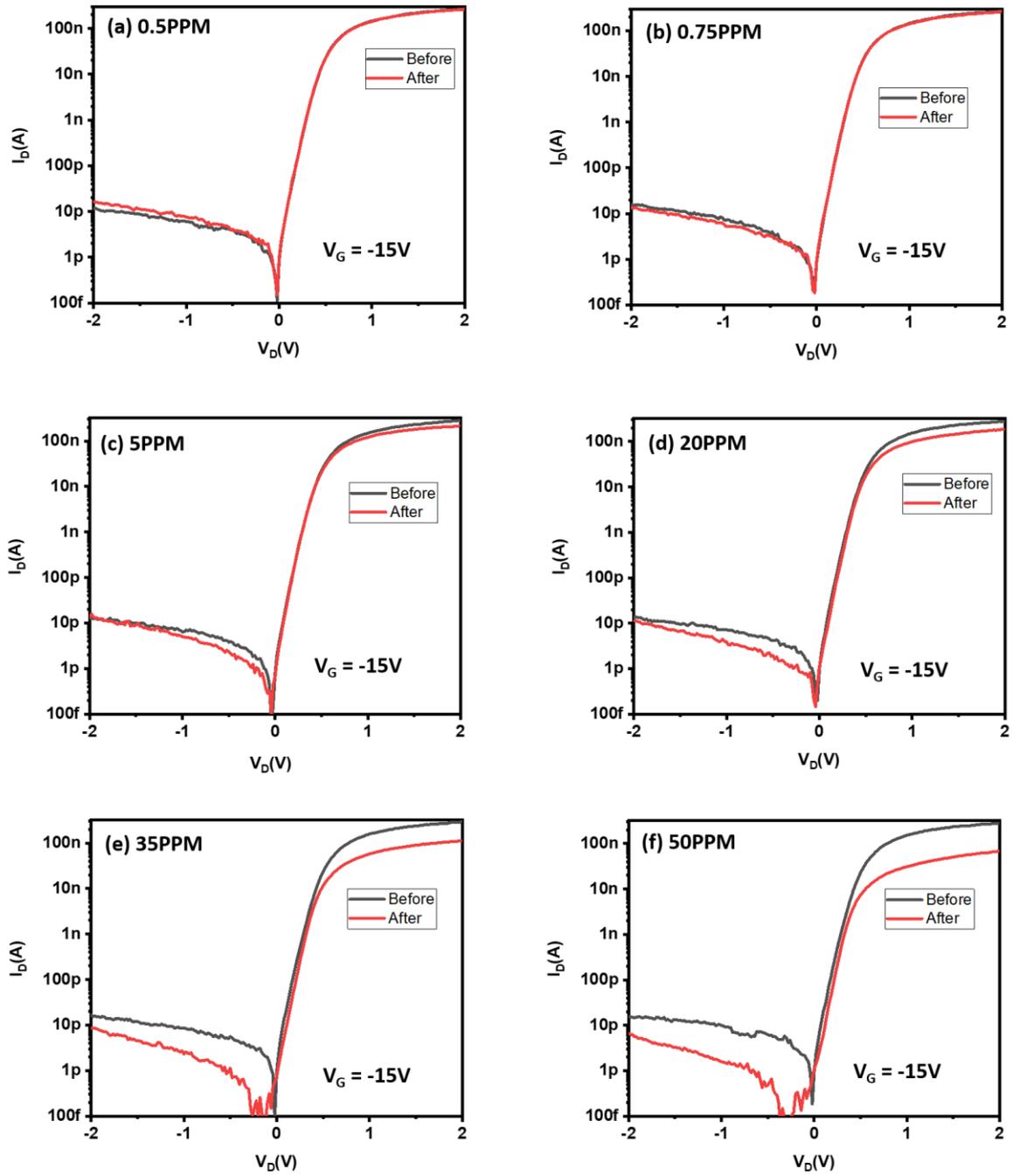

**Figure S4.** $I_D$-$V_D$ characteristics of the WSe$_2$/MoS$_2$ pn diode ($V_G$=-20V) before and after exposure to NO$_2$ gas at varying concentrations of (a) 0.5PPM, (b) 0.75PPM, (c) 5PPM, (d) 20PPM, (e) 35PPM, and, (f) 50PPM. The reduction in drain current increases with increasing NO$_2$ concentration, however the percentage decrease in current is higher for positive gate bias.

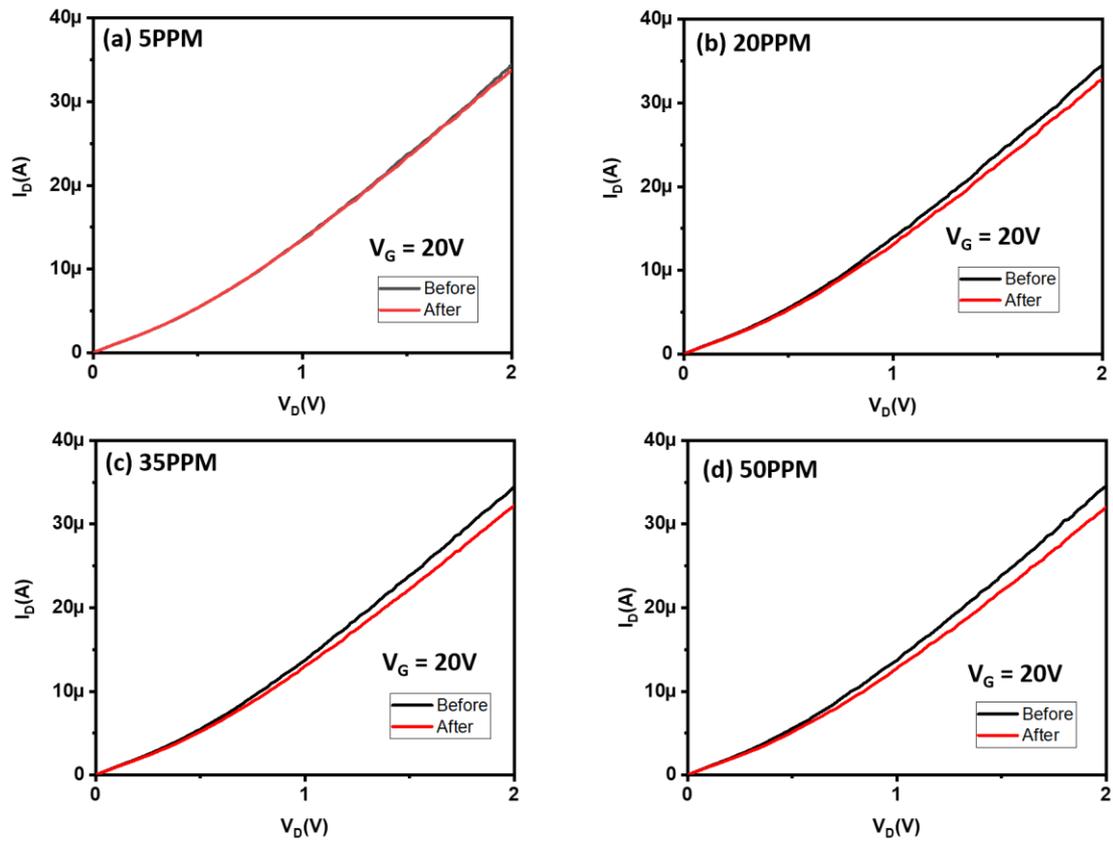

**Figure S5.** $I_D$-$V_D$ characteristics of the MoS$_2$ FET before and after exposure to NO$_2$ gas at varying concentrations of (a) 5PPM, (b) 20PPM, (c) 35PPM, and (d) 50PPM for +20V of gate biasing.

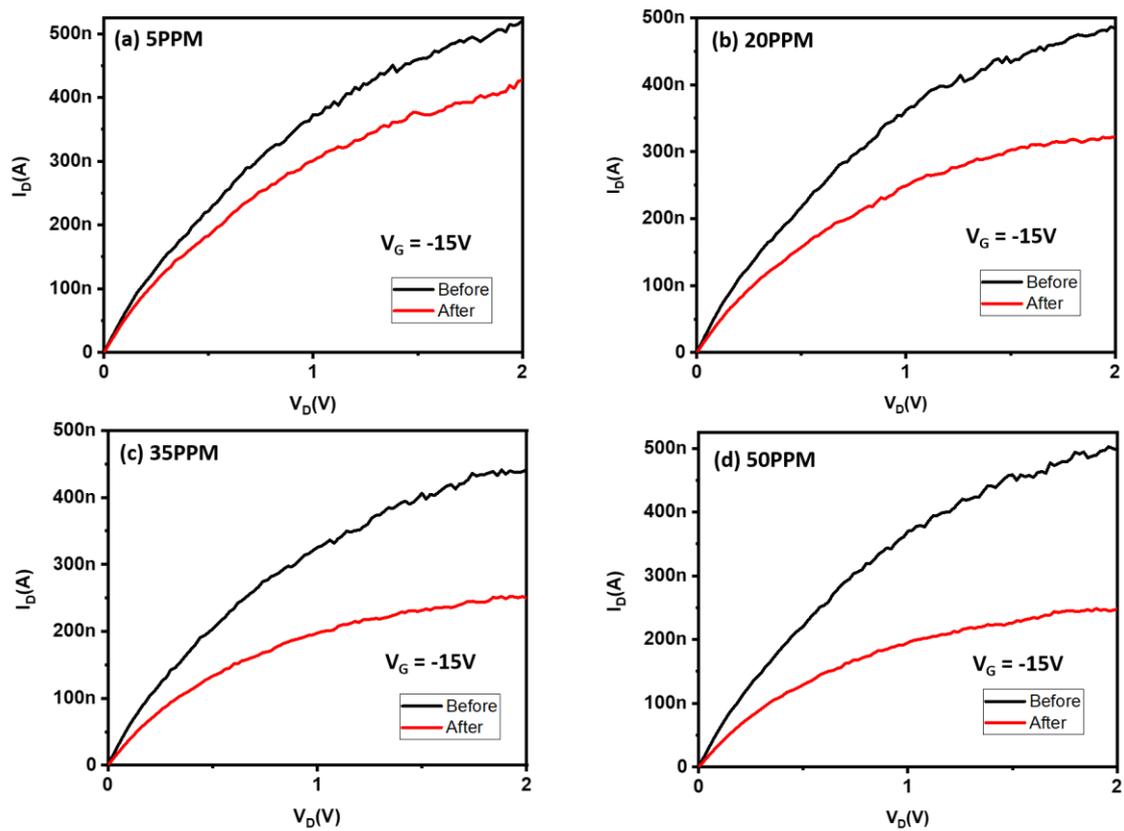

**Figure S6.** $I_D$-$V_D$ characteristics of the MoS$_2$ FET before and after exposure to NO$_2$ gas at varying concentrations of (a) 5PPM, (b) 20PPM, (c) 35PPM, and (d) 50PPM for -15V of gate biasing.

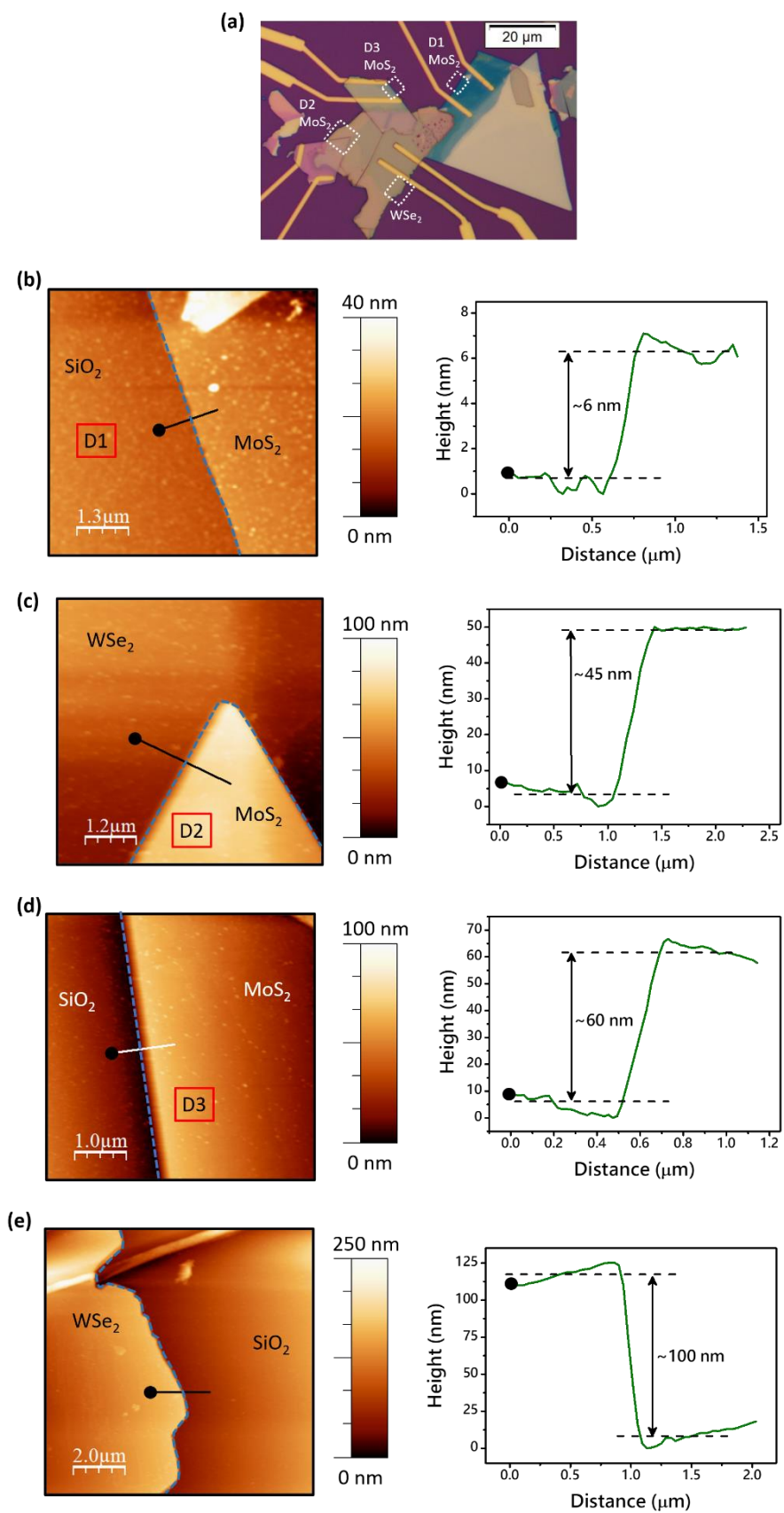

**Figure S7.** (a) Optical micrograph of the WSe$_2$/MoS$_2$ heterojunctions scanned to obtain AFM images and linescans of (b) thin, (c) medium and (d) thick MoS$_2$ flakes on a single, (e) thick WSe$_2$ flake.

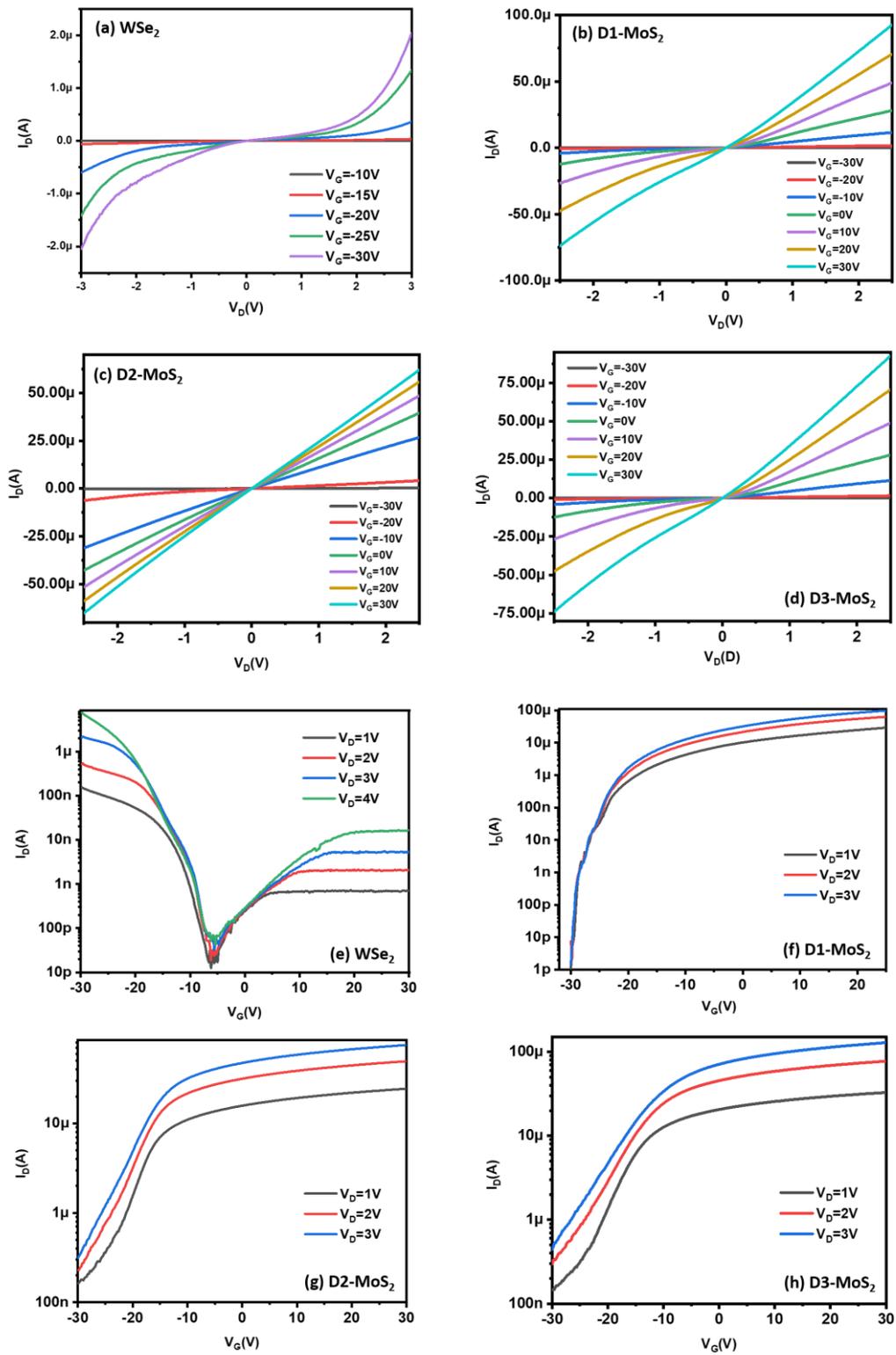

**Figure S8.** $I_D$-$V_D$ output characteristics of (a) WSe$_2$ and (b) D1, (c) D2, (d) D3 MoS$_2$ FETs. $I_D$-$V_G$ transfer characteristics of the (e) WSe$_2$ FET confirm its ambipolar nature. $I_D$-$V_G$ transfer characteristics of the (f) D1, (g) D2, (h) D3 MoS$_2$ FETs indicate n-type behavior.

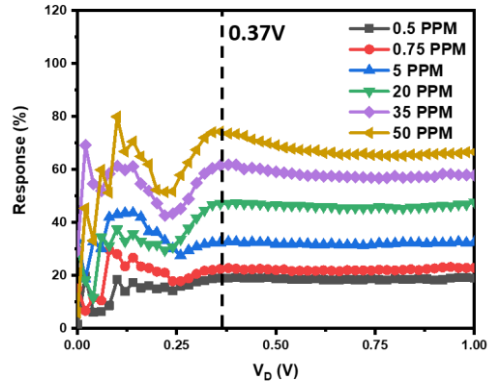

**Figure S9.** Drain biasing dependent sensing response of WSe$_2$/MoS$_2$ pn diode at +20V of gate bias.

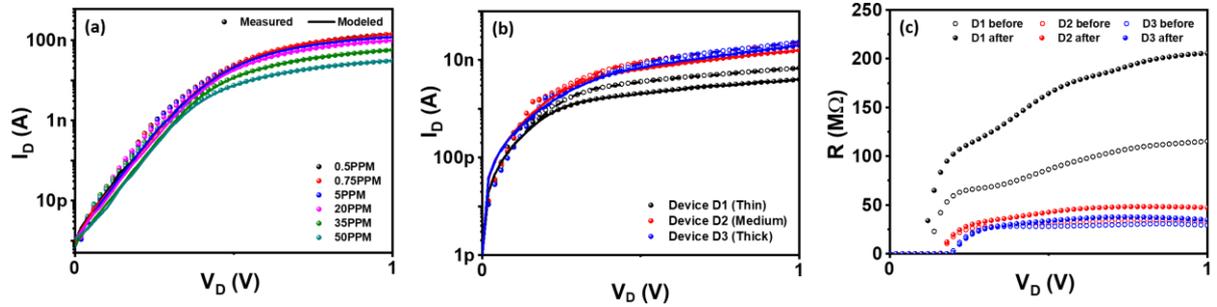

**Figure S10**. Equation $I = I_0(exp\left(\frac{V_D - IR}{\eta kT}\right) - 1)$ was used to model the WSe$_2$/MoS$_2$ pn diode characteristics. $I_0$ was extracted from $logI = logI_0 + \frac{V_D}{\eta kT}$ and R from $R = \frac{1}{I}\left[V - \eta kT \, log\left(\frac{I}{I_0} + 1\right)\right]$. (a) Forward bias model fits of WSe$_2$/MoS$_2$ pn diode characteristics for varying NO$_2$ concentration. Here $\eta$ is considered to be constant and evaluated from the slope of $logI$ vs $V_D$ plot at $V_D = 0.18V$. (b) Forward bias model fits of WSe$_2$/MoS$_2$ pn diode characteristics for varying MoS$_2$ flake thicknesses. (c) Extracted forward bias series resistance (R) before and after NO$_2$ exposure for diodes with varying MoS$_2$ flake thicknesses.

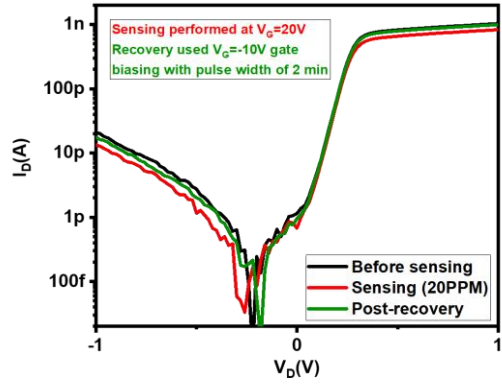

**Figure S11.** Complete device recovery shown through $I_D$-$V_D$ output characteristics measured before sensing, after exposure to 20PPM $NO_2$ and post-recovery after applying -10V of gate voltage for 2 min.

**Table S1.** Data used for limit of detection (LOD) calculations for the $WSe_2$/$MoS_2$ pn diode and the $MoS_2$ FET.

| Device | Biasing | Noise ($\sigma$) | Calibration curve slope (s) | LOD ($3\sigma/s$) |
|---|---|---|---|---|
| Diode | 1V | 0.39 | 8.19 | 0.15 |
| FET | 1V | 0.54 | 1.42 | 1.14 |